# (42355) Typhon-Echidna:
# Scheduling Observations for Binary Orbit Determination

W.M. Grundy[1], K.S. Noll[2], J. Virtanen[3], K. Muinonen[4], S.D. Kern[2],
D.C. Stephens[5], J.A. Stansberry[6], H.F. Levison[7], and J.R. Spencer [7].

1. Lowell Observatory, 1400 W. Mars Hill Rd., Flagstaff AZ 86001.
2. Space Telescope Science Institute, 3700 San Martin Dr., Baltimore MD 21218.
3. Finnish Geodetic Institute, Geodeetinrinne 2 (P.O.Box 15), FI-02431 Masala, Finland.
4. Dept. of Astronomy, Kopernikuksentie 1 (PO Box 14), FI-00014 University of Helsinki, Finland.
5. Formerly at Dept. of Physics and Astronomy, Johns Hopkins University, Baltimore MD 21218;
   now at Dept. of Physics and Astronomy, Brigham Young University, N283 ESC Provo UT 84602.
6. Steward Observatory, University of Arizona, 933 N. Cherry Ave., Tucson AZ 85721.
7. Southwest Research Institute, 1050 Walnut St. #400, Boulder CO 80302.

**ABSTRACT**

We describe a strategy for scheduling astrometric observations to minimize the number required to determine the mutual orbits of binary transneptunian systems. The method is illustrated by application to Hubble Space Telescope observations of (42355) Typhon-Echidna, revealing that Typhon and Echidna orbit one another with a period of 18.971 ± 0.006 days and a semimajor axis of 1628 ± 29 km, implying a system mass of (9.49 ± 0.52) × $10^{17}$ kg. The eccentricity of the orbit is 0.526 ± 0.015. Combined with a radiometric size determined from Spitzer Space Telescope data and the assumption that Typhon and Echidna both have the same albedo, we estimate that their radii are $76^{+14}_{-16}$ and $42^{+8}_{-9}$ km, respectively. These numbers give an average bulk density of only $0.44^{+0.44}_{-0.17}$ g cm$^{-3}$, consistent with very low bulk densities recently reported for two other small transneptunian binaries.

# 1. Introduction

Over 50 transneptunian objects (TNOs) have been discovered to be binaries in recent years, offering a rich source of information about the physical properties of this distant population (e.g., Noll et al. 2008). Transneptunian binaries (TNBs) also offer insight into the dynamical environments various sub-groupings of TNOs have experienced, by means of statistical studies of their abundances relative to single objects and of the ensemble characteristics of their mutual orbits (e.g., Petit and Mousis 2004; Kern and Elliot 2006; Noll et al. 2008).

An image in which a binary is discovered enables rudimentary characterization of the system in terms of the instantaneous sky-plane separation and relative brightness of the two components, but more detailed investigations require knowledge of their mutual orbit. For Keplerian motion (i.e., two gravitationally bound point masses without external influences), seven independent parameters are needed to completely describe the orbit. A single observation provides only two independent astrometric constraints: the relative separation and position angle, or equivalently, relative positions in two Cartesian axes. In principle, four observations could provide eight independent constraints, enough to completely determine the orbit, but in practice, constraints from separate observations often are insufficiently independent of one another. For example, four observations which happen to sample similar orbital longitudes are much less constraining of the geometry of the orbit than four observations well separated in orbital longitude. Since the period and geometry of the orbit are unknown *a priori*, there is no way of scheduling four observations in advance such that they reliably sample the orbit.

An additional challenge is that facilities able to do the observations necessary for TNB orbit determination are scarce. As seen from the Earth's location, these objects are extremely faint, typically $V > 22$ mag, and the components have very small separations, typically < 200 mas. Accordingly, the Hubble Space Telescope (HST), with its exceptionally stable, diffraction-limited images and its extremely low sky background, is the premier facility for their study. Large aperture, ground-based telescopes equipped with laser guide star adaptive optics can also contribute useful data. However, time on all such facilities is highly oversubscribed, and their various scheduling systems impose additional temporal constraints. It is essential to make the most efficient possible use of such valuable resources.

In this paper, we describe a scheme based on Monte Carlo techniques pioneered by Muinonen and Bowell (1993), Muinonen et al. (2001), and Virtanen et al. (2001, 2003, 2008), to use information from previous observations in scheduling subsequent observations to maximize their benefit for binary orbit determination. Using this method, we can typically find the period $P$, semimajor axis $a$, and eccentricity $e$ of an orbit within five observations (including the discovery observation). The remaining orbital elements (inclination $i$, mean longitude $\epsilon$ at a reference epoch, longitude of ascending node $\Omega$, and longitude of periapsis $\varpi$) can be secured with one or two additional observations, timed to take advantage of parallax from relative motion of the Earth and the target system. Our implementation was specifically designed with the scheduling constraints of HST in mind, but it could be adapted to mesh with scheduling modalities of other facilities.

To illustrate our methodology, we will show its application to the binary system (42355)

Typhon-Echidna[1], the first reported binary Centaur, with heliocentric osculating orbital elements $a_\odot$ = 37.9 AU, $e_\odot$ = 0.538, and $i_\odot$ = 2.43°. The binary nature of this system was discovered in 2006 January (Noll et al. 2006) from observations acquired through Noll et al. HST Cycle 14 program 10514 (hereafter "Visit 1"). Details of the clear filter combination and dithering techniques used in that program are described by Noll et al. (2006). Four subsequent follow-up observations (hereafter "Visits 2-5") were obtained during 2006 February, November, and December as part of Grundy et al. HST Cycle 14 program 10508. Observing parameters for that program are described by Grundy et al. (2007).

The scheduling strategy and scientific results of the follow-up observations are the subject of this paper.

## 2. Scheduling Follow-up Observations

Following the discovery of a TNB, we schedule follow-up observations in three distinct stages. In the first stage, we collect a pair of observations, preferably separated by less than half of one orbit. Since it typically takes about two weeks to activate and schedule an HST observation, it is often not possible to use the discovery observation as one of this pair, at least for the closer TNBs which have orbital periods of the order of days to weeks. Accordingly, we request a pair of observations separated by a time interval computed from the separation at discovery and the observed photometry, to be scheduled as soon as feasible. After those observations have been returned, we use a Statistical Ranging technique (Muinonen et al. 2001; Virtanen et al. 2001, 2003, 2008) to compute optimal timing for subsequent follow-up observations, subject to telescope scheduling constraints. This process is repeated until the period, semimajor axis, and eccentricity of the orbit are determined. Finally, one or two observations a year or more later are sometimes needed to resolve ambiguities between the angular elements of the orbit, by taking advantage of the relative motion of the Earth and the binary pair.

### 2.1 Phase One: Scheduling an initial pair

Our first task following discovery of a new binary is to schedule a pair of observations that are sufficiently separated in time to show orbital motion, but preferably less than half of one orbit. These observations are needed as inputs to the second phase to be described in Section 2.2. The period is not yet known, but the initial binary discovery images can be used to put some con-straints on it. Two relevant pieces of information are available from that observation: the instantaneous separation projected on the sky plane $r_{sky}$, and the instantaneous photometric brightnesses of the two components.

---

1  The names are from classical Greek mythology. The primary is named for Typhon (or Typhoeus), a flaming monster with 100 heads, the last offspring of Gaia. He was eventually vanquished by Zeus. The secondary is named for Echidna, with the face and torso of a woman and the body of a serpent. Typhon and Echidna produced numerous monstrous offspring including Cerberus (the 3-headed hound of Hades), the Chimaera, the Sphinx, and the Harpies. Their provisional designations were 2002 CR$_{46}$ and S/2006 (42355) 1. Typhon and Echidna were the first objects to be named according to a new naming convention adopted by the Committee on Small Body Nomenclature of the International Astronomical Union. Objects on unstable, non-resonant, giant-planet-crossing orbits with semimajor axes greater than Neptune's are to be named for hybrid mythical creatures other than Centaurs. Some 30 objects are currently known to fall into this category. So far, only one other object (the binary (65489) Ceto-Phorcys) has been named according to the policy, which acknowledges the dynamical similarity between these objects and objects on unstable, non-resonant, giant-planet-crossing orbits with semimajor axes smaller than 30 AU. Indeed, in the Deep Ecliptic Survey system of dynamical classification, Centaurs include both groups (e.g., Elliot et al. 2005; Noll et al. 2006).

The photometric brightnesses enable us to estimate the radii $R$ of the two components. For each component

$$R = \frac{664.5 \text{ km}}{\sqrt{A_p}} 10^{\frac{-H_V}{5}}, \tag{1}$$

where $A_p$ is an assumed V-band geometric albedo and $H_V$ is the absolute V magnitude of the body (e.g., Bowell et al. 1989; if data on phase behavior are unavailable, we assume $G = 0.15$ in the $H$ and $G$ system to convert a V magnitude to $H_V$). For initial observations using the less well calibrated ACS/HRC clear filter combination, V photometry from the literature is preferable when it is available. Clear filter images can, however, provide the photometric difference between primary and secondary. For Typhon and Echidna, the Visit 1 clear filter magnitude difference was reported as 1.47 ± 0.04 (Noll et al. 2006) and the combined $H_V$ as 7.65 ± 0.01 (Tegler et al. 2003), from which we estimated Typhon's $H_V$ as 7.90 and Echidna's as 9.37. We assumed albedos of 0.03 and 0.25 bracketed the range of probable values, with an expected value around 0.08 (e.g., Stansberry et al. 2008). Assuming a range of densities (we used 0.5 and 2.0 g cm$^{-3}$ to define the range, with 1 g cm$^{-3}$ as an expected value) we converted the range of radii into a range of masses for each component. Summing the component masses gave a plausible range of system masses $M_{sys}$ spanning about two orders of magnitude. For the Typhon-Echidna system we estimated that $M_{sys}$ should be between $10^{17}$ and $10^{19}$ kg, with an expected value around $10^{18}$ kg.

The separation on the sky plane at the time of discovery $r_{sky}$ offers some constraint on the semimajor axis $a$ of the orbit. For a bound orbit, it is impossible for $a$ to be less than one half of $r_{sky}$. Likewise, $a$ is unlikely to be many times larger than $r_{sky}$, since that would require improbable orbit orientation as well as improbable timing of the observation. To be more quantitative, we generated orbits at random orientations with random eccentricities, and sampled them at random times to find the distribution of $a$ relative to the observed $r_{sky}$, as shown in Fig. 1. The mode, the most probable value of $a$, is near $r_{sky}$. The mode and the width of the distribution are relatively insensitive to the assumed probability distribution of eccentricities $e$. For $e$ uniformly distributed between 0.0 and 0.8 (the range for known TNB mutual orbits, e.g., Noll et al. 2008), the mode is $1.01 \times r_{sky}$ and 68% of $a$ are within $\pm 0.43 \times r_{sky}$ of the modal value. For Typhon-Echidna, $r_{sky}$ at Visit 1 was 1330 ± 130 km (Noll et al. 2006)., leading us to expect $a$ of 1340 km, with a plausible range of 700 to 2100 km.

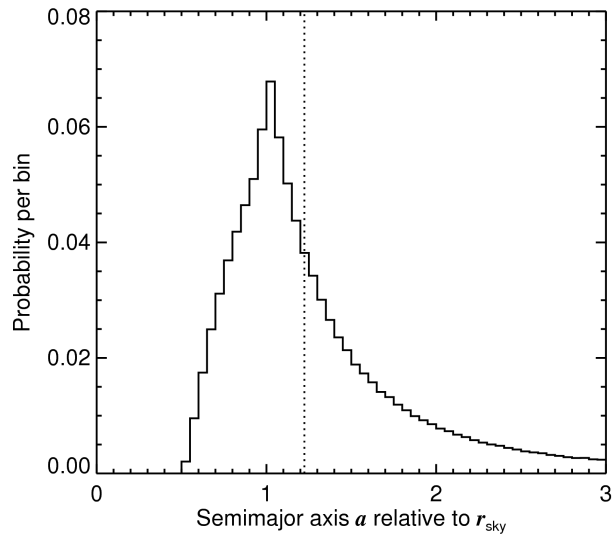

Fig 1. Distribution of semimajor axes $a$ in terms of initial discovery sky plane separation $r_{sky}$, for a collection of random orbits sampled at random times. We assumed uniformly distributed random eccentricities $e$ between 0.0 and 0.8, although other plausible $e$ distributions give similar results. The most probable value of $a$ is very close to $r_{sky}$ and $a$ is within a factor of 2 of $r_{sky}$ 84% of the time. The vertical dotted line shows the actual ratio of $a/r_{sky}$ for Typhon and Echidna, determined later in this paper.

Combining the plausible $M_{sys}$ and $a$ ranges we obtain a range of plausible orbital periods $P$, via the relation

$$P = 2\pi \sqrt{\frac{a^3}{G M_{sys}}}, \qquad (2)$$

where $G$ is the gravitational constant, which we take to be $6.6742 \times 10^{-11}$ m$^3$ s$^{-2}$ kg$^{-1}$ (CODATA 2006). For the Typhon-Echidna system, this calculation gave a plausible period range of 2 to 85 days, with an expected period of 14 days.

Our goal is to obtain a pair of observations sufficiently separated in time to reveal significant orbital motion, preferably separated by less than half of an orbit for reasons to be explained in Section 2.2. To accomplish this, we request a pair of observations separated by a quarter of the expected period, but include a 1-2 day tolerance, since HST scheduling is constrained by many additional factors outside of our control. For this system, the two follow-up observations (Visits 2 and 3) ended up being separated by about 4.5 days.

## 2.2 Phase Two: Scheduling subsequent observations

Combining the discovery observation with the two follow-up observations, we have enough data to begin to significantly constrain the universe of possible Keplerian orbits. To characterize that population of possibilities, we employ a method adapted from the Statistical Ranging technique for determination of heliocentric orbits. The concept is simple: generate numerous random orbits, keeping only those which are consistent with the observations. The resulting collection of orbits can provide an approximation of the probability density function (PDF) in 7-dimensional Keplerian orbital element space. It can be used to explore which regions of orbital element space are most consistent with the observations, and can also be used to optimize scheduling of subsequent observations based on how the collection of sky-plane positions of the orbits evolves over time. Observations at times when that cloud of sky plane positions is most diffuse are most useful for whittling away at the ensemble of remaining possible orbits.

The primary difficulty is that prohibitively few truly random orbits will actually satisfy the observational data, so it is essential to start with random orbits that already satisfy one or more of the data points. In the Virtanen et al. Statistical Ranging approach to heliocentric orbit determination, random distances are generated along two of the observed lines of sight[2], which we will call the "anchor" observations. These two points in space, combined with the anchor observation times, specify a unique prograde heliocentric orbit (which may or may not be bound). Sky-plane positional predictions of this orbit can be tested against other available observations by means of the $\chi^2$ statistic, and the orbit rejected if they are inconsistent.

---

[2] Other methods exist. For example, Hestroffer et al. (2005) employ the method of Thiele-Innes, based on 3 anchor observations instead of 2. That method and our method showed similar results when applied to the Typhon-Echidna astrometry (D. Hestroffer, personal communication 2006). Margot et al. (2005) have also developed a Monte Carlo method for this purpose, but the details have not been published.

To adapt the Statistical Ranging technique to TNB orbit determination requires a few modifications. Most importantly, unlike the heliocentric case, we do not know the system mass $M_{sys}$. Without this piece of information, we need to generate random masses in addition to random distances along the two anchor lines of sight (alternatively, one could generate random periods). We use a uniform distribution in $\log(M_{sys})$ spanning the range of plausible masses from Section 2.1. Another adjustment relates to the range of possible random locations along the anchor lines of sight. Most likely the secondary was within a factor of a few times $r_{sky}$ of the primary at the times of the anchor observations, so it is unnecessary to generate random positions far outside that range. We draw random numbers from a Gaussian probability distribution centered on the radial distance of the primary, with a 1-$\sigma$ width equal to three times the maximum observed sky plane separation. To account for the astrometric uncertainties of the anchor observations, the same is done on the sky plane, using Gaussian probability distributions centered on the anchor points with 1-$\sigma$ widths set by the astrometric error bars.

Obtaining orbital elements from the two anchor locations and times plus a system mass is a two-point boundary value problem which can be solved in a number of ways. Virtanen et al. (2001) used the method of successive fractions (e.g., Dubyago 1961; Bate et al. 1971), but this method does not work for situations where the secondary has moved through a large angle relative to the primary between the times of the two anchor points. Instead we use the "$p$-iteration" method (also known as the method of Herrick and Liu) which does not suffer from this particular limitation (e.g., Danby 1992; Granvik and Muinonen 2005; Granvik 2008). There are two potential solutions using this method, one for the secondary going around the shorter way between the two anchor points, and the other going around the longer way. Going around the longer way implies more rapid motion, which can sometimes be excluded by examining multiple images taken over the course of a single visit, but if the direction of motion of the secondary is ambiguous, we are obliged to try both solutions (or to choose one of the two at random), doubling the computational cost. Accordingly, we prefer anchor observations separated by much less than half of an orbit, so that the direction of motion is obvious.

A weight is calculated for each orbit as described by Virtanen et al (2001, 2008) and by Virtanen and Muinonen (2006). These weights account for the goodness of agreement to the observational data and for our Monte Carlo sampling in range and sky plane positions rather than in the orbital elements themselves. Factors of the reciprocal of the probability of having selected each of the two random anchor positions are included to correct for the non-uniform spatial sampling. The weights are normalized such that their sum is unity. They indicate how much each orbit contributes to the representation of the PDF in orbital element space. The structure of the PDF tends to be dominated by a small fraction of random orbits having the highest weights. The majority have weights far too small to contribute appreciably to the PDF and can safely be discarded (see Virtanen et al. 2008). High-weighted orbits are collected until enough are available for further analysis. How many are enough? Rigorous criteria are discussed by Virtanen et al. (2008), but in practice, we find that a smaller sample usually characterizes the PDF sufficiently well for our purposes when the most highly-weighted orbit contributes 10% or less of the total and scatterplots of half of the orbit collection look like scatterplots of the other half. Figure 2 shows scatterplot representations of the orbital element PDF for Echidna's orbit around Typhon, based on the three initial HST observations, using Visits 2 and 3 as anchors. Points in the figure represent Monte Carlo orbits consistent with the three observations, scaled according to their weights.

To schedule the next observation, we make use of this knowledge of the PDF. Consider the sky plane positions computed from the collection of orbits as a function of time (with weights adjusted to account for the transformation from orbital elements to spatial coordinates). A follow-up observation will be most useful if it is obtained when this cloud of positions extends over the largest possible region on the sky plane, since an observation at that time can exclude the largest region of orbital element space. This sky plane cloud is shown for two example dates in Fig. 3.

The extent of the sky plane cloud is a somewhat nebulous concept for a cloud having a complicated shape. It could be quantified in a variety of ways ranging from the standard deviation to the area within a convex hull. We use the mean distance between any two points in the cloud, weighting the distance between each pair according to the square root of the product of the weights of the two associated orbits. This measure, shown in Fig. 4, indicates the astrometric precision needed to distinguish any one candidate orbit from another, on average. From this example, we would have requested an observation during 2006 February 21-29, March 15-20, or April 5-10, if any of those dates could have been scheduled, as indicated by horizontal bars.

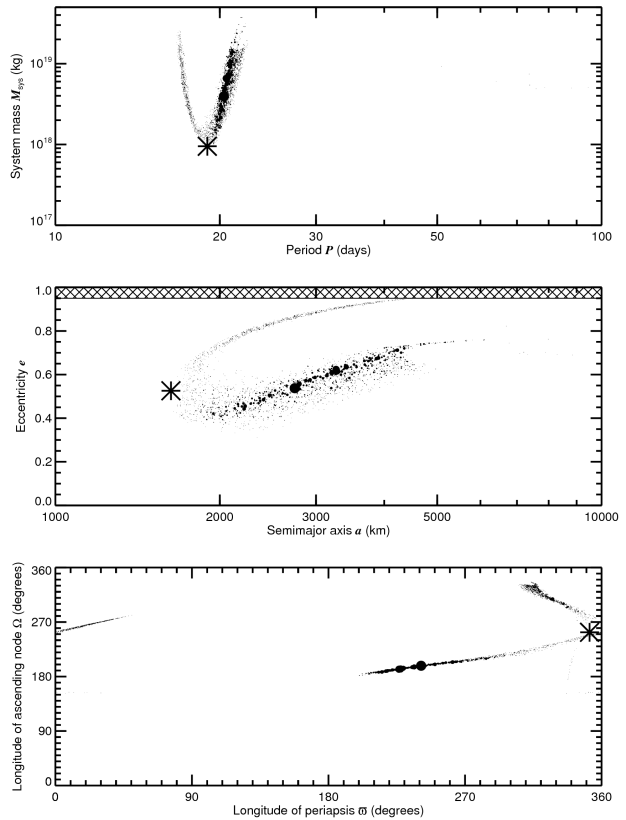

Fig. 2: Each point in these scatterplots represents a randomly-generated orbit consistent with astrometry of Echidna relative to Typhon from HST Visits 1-3. Points are sized in proportion to their weights. By default, the hatched area in the middle panel ($e > 0.95$) is not explored by our code to save CPU time. Orbits are grouped into a few broad swathes, with large areas of orbital element space essentially unoccupied. The actual values, determined later in this paper, are indicated by asterisks.

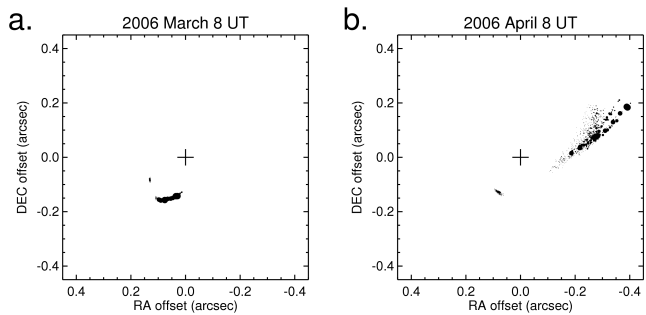

Fig. 3: Locations of the secondary relative to the primary, projected onto the sky plane for an Earth-based observer for two dates, as predicted by the cloud of Monte Carlo orbits based on Visits 1-3 and shown in Figure 2. Each point represents one Monte Carlo orbit, weighted as described in the text. More of orbital element space could be ruled out by an observation done at the time of the right panel.

During 2006, HST was operated using only two instead of the usual three gyroscopes in order to prolong the life of its remaining gyroscopes. One consequence of 2-gyro mode is diminished observability windows on the sky. The Typhon-Echidna system became unobservable to HST for a period of many months beginning shortly after Visit 3. By the time it became observable again in November, the cloud of potential sky-plane positions was so homogenized that there were no discernible local maxima, so we simply requested Visit 4 take place at the earliest available opportunity, which ended up being 2006 November 5 UT. Combining Visit 4 with the previous three, the allowed orbital element space became much more constrained, as shown in Figs. 5 and 6. Based on this new set of Monte Carlo orbits, we requested Visit 5 to take place during December 3-6 UT. Unfortunately, guide star lock failed during that observation and no useful data were obtained, so we rescheduled Visit 5 for the next period of maximum sky-plane cloud extent, December 22-25 UT. That observation was successful, resulting in the final astrometric data set in Table 1.

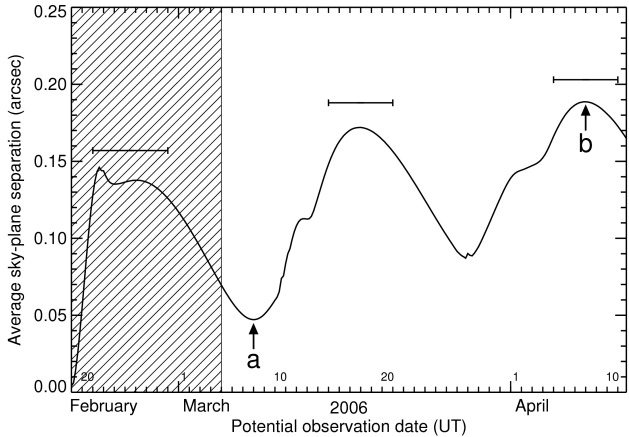

Fig. 4: The weighted average separation of sky-plane positions for the Monte Carlo orbit collection as a function of time following Visit 3 (2006 February 19 UT). Local maxima around February 21-29, March 15-20, and April 5-10 indicated by horizontal bars would be better dates for a follow-up observation. Times of the two panels shown in Fig. 3 are indicated with arrows. Unfortunately, none of those dates could be scheduled. The hatched region was unobservable because of HST scheduling lead time constraints and dates after about March 1 were excluded as a result of 2-gyro mode restrictions.

Table 1. Hubble Space Telescope ACS/HRC observational circumstances and relative astrometry

| Visit number | Average UT date and hour | $r$ [a] | $\Delta$ [a] | $g$ [a] | $\Delta x$ [b] | $\Delta y$ [b] |
|---|---|---|---|---|---|---|
| | | (AU) | | (degrees) | (arcsec) | |
| 1 | 2006/01/20 10.03 | 17.529 | 16.675 | 1.64 | −0.08727 | −0.08172 |
| 2 | 2006/02/14 13.25 | 17.528 | 16.543 | 0.24 | +0.06642 | −0.15552 |
| 3 | 2006/02/19  1.74 | 17.528 | 16.539 | 0.07 | +0.13570 | −0.08515 |
| 4 | 2006/11/05  1.33 | 17.533 | 17.920 | 2.95 | −0.00355 | −0.13523 |
| 5 | 2006/12/22 10.85 | 17.538 | 17.135 | 2.97 | +0.10178 | +0.00323 |

[a.] The distance from the Sun to the target is $r$ and from the Earth to the target is $\Delta$. The phase angle, the angular separation between the Earth and Sun as seen from the target, is $g$.

[b.] Relative right ascension $\Delta x$ and relative declination $\Delta y$ are computed as $\Delta x = (\alpha_2 − \alpha_1)\cos(\delta_1)$ and $\Delta y = \delta_2 − \delta_1$, where $\alpha$ is right ascencion, $\delta$ is declination, and subscripts 1 and 2 refer to Typhon and Echidna, respectively. 1-$\sigma$ uncertainties are estimated as ± 2.9 mas for Visit 1 and ± 1.9 mas for Visits 2-5.

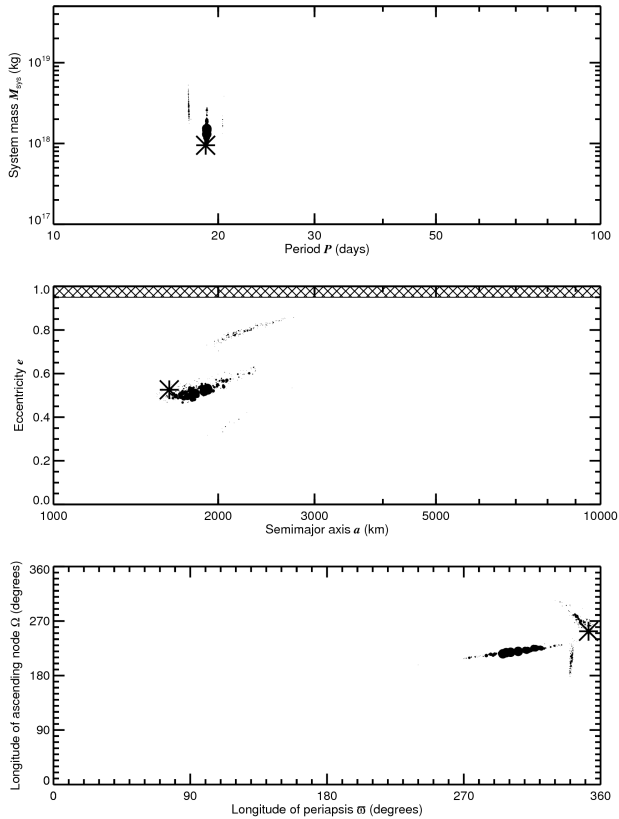

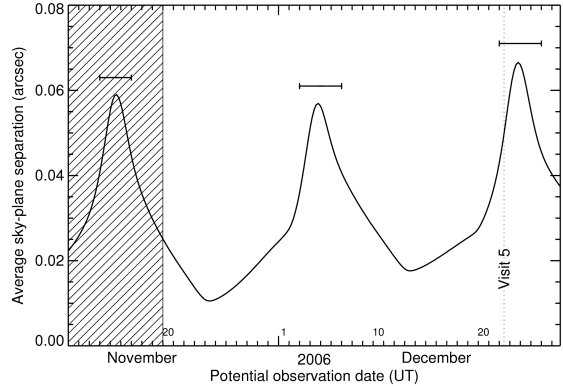

Fig. 5: Same as Fig. 2, but now including astrometry from HST Visits 1-4. Allowed periods and eccentricities have collapsed into one tight, primary clump along with two adjacent low-probability clumps. The primary clump includes the actual orbit, as indicated by the asterisks.

Fig. 6: Same as Fig. 4, except following HST Visit 4 2006 November 5 UT. The next available favorable observing opportunity according to these curves was December 3-6 UT. Visit 5 was attempted then, but failed. It was rescheduled for the next subsequent opportunity December 22-25 UT, and was executed at a mean time of 2006 December 22 10:51 UT, as indicated by the dotted line.

### 2.3 Breaking the mirror ambiguity

As observational constraints accumulate, the Monte Carlo orbits occupy progressively fewer and smaller clumps in orbital element space. Ultimately, each clump corresponds to an individual orbit solution. When the data permit only a single such clump, we have found the unique solution. Often, two "mirror" clumps remain, with common $P$, $a$, and $e$, but different $i$, $\epsilon$, $\Omega$, and $\varpi$. This happens when the geometry between the observer and the binary system is essentially unchanged over the course of the observations. Under these circumstances, the actual orbit is astrometrically indistinguishable from its reflection through the sky plane, as reviewed by Descamps (2005). For Solar System binaries, this mirror ambiguity can be broken by taking advantage of parallaxes arising from the relative motions of Earth and of the binary system along their heliocentric orbits. Distinguishing between the two mirror orbit solutions usually only requires one more strategically-timed observation. The timescale needed to distinguish between mirror orbits depends on several factors, including how far the binary is from the Sun, the size of the binary orbit, and its orientation. More nearby binaries, with larger and/or more edge-on orbits can generally be resolved sooner.

The Monte Carlo techniques of Section 2.2 could in principle be used to optimize schedul-ing of observations to break the mirror symmetry, but it becomes more computationally expensive as the volume of allowed orbital element space consistent with the available data shrinks. We find that when the remaining allowed orbits are confined to tight clumps, it becomes more efficient to directly fit Keplerian candidate orbits to the astrometric data as described in Grundy et al. (2007), using one of the Monte Carlo orbits from each clump as a starting vector. We use the "amoeba" downhill simplex algorithm (Nelder and Mead 1965; Press et al. 1992) to iteratively adjust the seven orbital elements ($P$, $a$, $e$, $i$, $\epsilon$, $\Omega$, and $\varpi$) to minimize the $\chi^2$ statistic between the observed and predicted astrometry, accounting for light time delays and changing geometry between the observer and binary system. The uncertainties associated with each candidate orbit can be quantified by fitting orbits to points randomized around the actual observations and error bars to rapidly generate a clump of random orbits surrounding each candidate orbit.

For the Typhon-Echidna mutual orbit, no additional observations were needed. The system is relatively close to the Sun and our observations spanned almost an entire year, resulting in sufficiently different viewing geometries to break the mirror ambiguity. The addition of the data from Visit 5 collapsed the Monte Carlo orbits to a single clump corresponding to the unique solution to be described in Section 3.

When additional observations are needed, an optimal observing time can be found by computing the separation between sky plane positions predicted by the candidate orbits as a function of time. To distinguish between the two requires a single observation at a time when the two sky-plane positions are separated by more than the obtainable astrometric precision. We express this criterion as a dimensionless "distinguishability" ratio, defined as the separation divided by precision, as shown in Fig. 7. It is desirable to observe when this ratio is as high as possible. It should be at least three, including the effect of the gradually expanding clouds of positions associated with the uncertainties of each orbit (otherwise an additional observation may be needed).

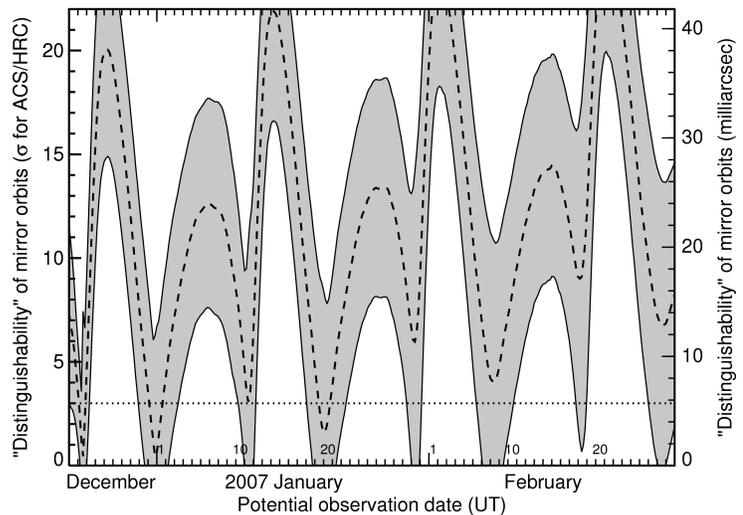

Fig. 7: Sky plane separation between positions predicted by the two best fit mirror orbits divided by achievable astrometric precision (1.9 mas for ACS/HRC) as a function of time, which we call the "distinguishability" of the two orbits (dashed curve). The gray area indicates the effect of uncertainties in the orbital elements. This plot was not needed, since the mirror ambiguity had already been broken by our five observations. However, if the orbit had not already been secured, one more observation at a time when the "distinguishability" was larger than the astrometric uncertainty would have clinched it. The 3-$\sigma$ threshold is indicated by a horizontal dotted line. The right axis is labeled for milliarcsec, to enable comparison with capabilities of other possible instruments.

## 3. Results for the Typhon-Echidna system

Final astrometric and photometric measurements are listed in Tables 1 and 2. Our reduction procedures have been slightly updated from methods described in Grundy et al. (2007), as follows. Individual HST ACS/HRC images (files with ".flt" appendages) are analyzed by iteratively fitting a pair of sub-sampled Tiny Tim PSFs (Krist and Hook 2004) to the observed HST image using the "amoeba" algorithm. The image background and focus parameters are fit independently. Jitter is approximated with Gaussian smoothing. Fitted astrometric positions are corrected for ACS/HRC field distortion and a pixel area map correction applied to account for projected pixel sizes. Photometry measured on the individual model PSFs is corrected to an infinite aperture using values from Sirianni et al. (2005; Table 5). Photometric zero points are determined from the PHOTFLAM and PHOTZPT values in the image header in combination with Synphot conversion from the STmag to Vegamag reference system (Laidler et al. 2007).

Our photometry shows similar colors for the two components: Typhon's mean $V\!-\!I$ color is $0.99 \pm 0.04$ mag and that of Echidna is $0.95 \pm 0.04$ mag. From these numbers, we derive a system average photometric color slope of $S = 11.1 \pm 1.3$ (% change per 100 nm relative to $V$), slightly less red than the $15.9 \pm 1.9$ average slope computed from photometric colors for this system reported by Peixinho et al. (2004) and by Tegler et al. (2003). Both color slopes put this object among the gray group of Centaurs (e.g., Peixinho et al. 2003, 2004; Tegler et al. 2003). Combining all available observations, Echidna is $1.30 \pm 0.06$ mag fainter than Typhon. Variations in the photometry suggest the possibility of lightcurves. After correcting for geometric effects (by assuming $G = 0.15$ in the $H$ and $G$ system of Bowell et al. 1989), Typhon was about 0.15 mag brighter than average during Visit 2 in both $F606W$ and $F814W$ filters. Likewise, Echidna was brighter than average during Visits 2 and 4 and fainter in Visits 3 and 5, in both filters. The absolute magnitude computed from our photometry is $H_V = 7.73 \pm 0.04$ mag, similar to the value of $7.65 \pm 0.01$ mag reported by Tegler et al. (2003).

Table 2. Photometry from Hubble Space Telescope ACS/HRC observations (magnitudes)

| Visit number | Observation Date (UT) | Typhon F606W | F814W | V | I | Echidna F606W | F814W | V | I |
|---|---|---|---|---|---|---|---|---|---|
| 2 | 2006/02/14 | 20.052 | 19.227 | 20.301 | 19.255 | 21.375 | 20.605 | 21.624 | 20.633 |
| 3 | 2006/02/19 | 20.114 | 19.349 | 20.364 | 19.377 | 21.474 | 20.777 | 21.724 | 20.805 |
| 4 | 2006/11/05 | 20.623 | 19.907 | 20.873 | 19.935 | 21.675 | 20.983 | 21.925 | 21.011 |
| 5 | 2006/12/22 | 20.552 | 19.774 | 20.801 | 19.802 | 21.779 | 21.013 | 22.028 | 21.041 |

Note: Photometric uncertainties of $\pm 24$ millimag (1-$\sigma$) were estimated from the scatter among four frames taken through each filter during each visit. Color transformation uncertainties inflate $V$ and $I$ uncertainties to around $\pm 30$ millimag.

Best fit orbital elements for Echidna relative to Typhon are tabulated in Table 3. The $\chi^2$ for this solution is 3.0 ($\chi^2$ for the mirror orbit solution is about 50 and is formally excluded). Sky-plane residuals for this solution range from 0.6 to 3.3 mas, with an average of 1.6 mas. The location of the orbit pole ($\alpha = 163 \pm 5$, $\delta = 52 \pm 2$ degrees J2000 equatorial) makes the orbit prograde. Interestingly, the plane of this orbit will sweep across the inner Solar System during the 2019-2026 time frame, providing numerous opportunities to observe mutual occultations and eclipses, the precise timing of which could lead to more accurate measurement of the orbit plane and of the combined size of the two bodies (e.g., Binzel and Hubbard 1997; Descamps et al. 2007; Noll et al. 2008). Potential precession effects were not considered in our orbit fits. If such effects were acting on this system, they could lead to period errors and errors in the timing of mutual events. More observations would be needed to test for such effects.

Table 3. Orbital parameters and 1-$\sigma$ uncertainties for Echidna relative to Typhon

| Parameter | | Value |
|---|---|---|
| Fitted elements:[a] | | |
|    Period (days) | $P$ | 18.9709 ± 0.0064 |
|    Semimajor axis (km) | $a$ | 1628 ± 29 |
|    Eccentricity | $e$ | 0.526 ± 0.015 |
|    Inclination[b] (deg) | $i$ | 37.9 ± 2.0 |
|    Mean longitude[b] at epoch[c] (deg) | $\epsilon$ | 11.5 ± 1.6 |
|    Longitude of asc. node[b] (deg) | $\Omega$ | 253.1 ± 4.0 |
|    Longitude of periapsis[b] (deg) | $\varpi$ | 352.1 ± 1.4 |
| Derived parameters: | | |
|    System mass ($10^{17}$ kg) | $M_{sys}$ | 9.49 ± 0.52 |
|    Orbit pole right ascension[b] (deg) | $\alpha_{pole}$ | 163.0 ± 4.5 |
|    Orbit pole declination[b] (deg) | $\delta_{pole}$ | 52.2 ± 2.3 |

[a.] Goodness of fit for this solution is $\chi^2 = 3.0$. The average sky plane residual is 1.6 mas. Uncertainties on fitted parameters were derived as described by Grundy et al. (2007).

[b.] Referenced to J2000 equatorial frame.

[c.] The epoch is Julian date 2454000.0 (2006 September 21 12:00 UT).

From the semimajor axis $a$ and period $P$, we can compute the mass of the combined system $M_{sys}$, by rewriting equation 2 as

$$M_{sys} = \frac{4\pi^2 a^3}{G P^2}. \tag{3}$$

For Typhon-Echidna, we obtain $M_{sys} = (9.49 \pm 0.52) \times 10^{17}$ kg. The semimajor axis is the dominant source of uncertainty in $M_{sys}$.

Spitzer Space Telescope observed the Typhon-Echidna system during a single visit centered on 2004 April 13 at 3:44 UT, using the 24 and 70 μm channels of its Multiband Imaging Photometer for Spitzer (MIPS) instrument (Rieke et al. 2004). These observations were part of Stansberry et al. program 55. The system was not detected at 24 μm, but was detected with a signal to noise ratio of about 9 at 70 μm, enabling Stansberry et al. (2008) to estimate a geometric albedo $A_p = 0.0509^{+0.0124}_{-0.0080}$ and an effective radius $R_{eff} = 86.9^{+7.8}_{-9.0}$ km (the radius of a sphere having the same total projected surface area). These results are somewhat more speculative than other Spitzer TNO radiometric sizes and albedos since only one thermal wavelength was detected, necessitating assumptions about the thermophysical behavior of the surfaces of the bodies. To ap-proximately account for this sensitivity, as well as the possibility that the single observation could have sampled a lightcurve peak or valley, we used twice these reported uncertainties. Dividing the surface area implied by $R_{eff}$ between Typhon and Echidna according to their relative brightnesses (1.30 ± 0.06 mag) results in radii of $76^{+14}_{-16}$ and $42^{+8}_{-9}$ km, respectively, assuming equal albedos for the two (a reasonable assumption, in light of their similar colors). Computing the total system volume from these radii, and combining that with the system mass from the mutual orbit, we obtain an average bulk density of $0.44^{+0.44}_{-0.17}$ g cm$^{-3}$.

The density of $0.44^{+0.44}_{-0.17}$ g cm$^{-3}$ is remarkably low, about half that of water ice, but it is not without precedent among small TNBs. Stansberry et al. (2006) reported a density of $0.50^{+0.30}_{-0.20}$ g cm$^{-3}$ for (47171) 1999 TC$_{36}$ and Spencer et al. (2006) reported a density of $0.70^{+0.32}_{-0.21}$ g cm$^{-3}$ for (26308) 1998 SM$_{165}$. Low densities have also recently been reported for the Trojan asteroid (617) Patroclus (Marchis et al. 2006) and for much smaller comet nuclei (e.g., A'Hearn et al. 2005; Davidsson and Gutiérrez 2006). Several much larger, planet-sized TNOs have also had their bulk densities measured, including Pluto and Charon (2.03 ± 0.06 and 1.65 ± 0.06 g cm$^{-3}$, respectively; Buie et al. 2006), Eris (2.26 ± 0.25 g cm$^{-3}$; Brown 2006), and (136108) 2003 EL$_{61}$ (3.0 ± 0.4 g cm$^{-3}$; Rabinowitz et al. 2006). This modest sample of TNO densities is plotted versus primary radius in Figure 8. There seems to be a transition between smaller objects with lower densities and larger objects with higher densities somewhere between radii of 200 and 600 km that is greater than that predicted by self compaction models (e.g., Lupo and Lewis 1979; McKinnon et al. 2005). Similar arguments have been made by Lacerda and Jewitt (2007) based on densities from rotating fluid body models of high amplitude lightcurves. The existence of such a contrast suggests distinct thermal histories or even compositional differences between the two groups of TNOs, although the relatively high density of the small system (65489) Ceto-Phorcys ($1.37^{+0.66}_{-0.32}$ g cm$^{-3}$; Grundy et al. 2007) injects a cautionary note and points to the need to bolster the meager sample of estimated TNO densities.

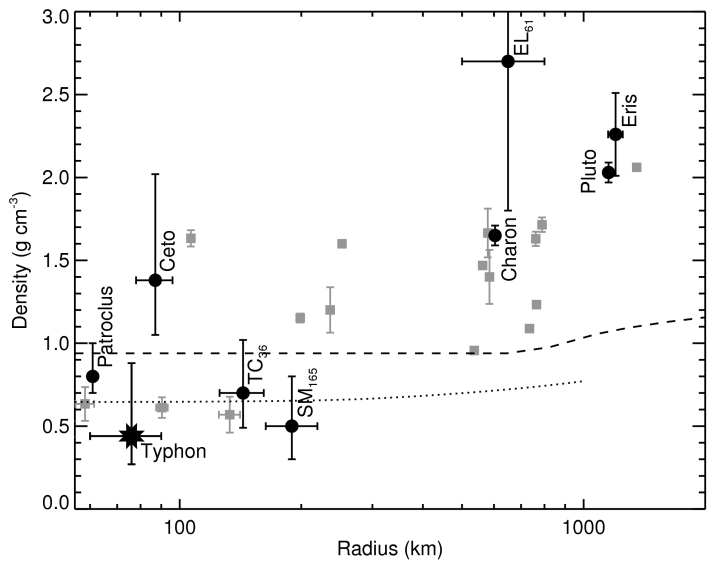

Fig. 8: Comparison of the average bulk density and primary radius of Typhon-Echidna (star) with densities and primary radii of other outer Solar System binaries mentioned in the text (labeled black circles) and of icy satellites of Saturn, Uranus, and Neptune (gray squares, from Burns 1986 and McKinnon et al. 1995). The dashed curve is a theoretical bulk density for a non-porous, pure H$_2$O ice sphere subject to self-compression at 77 K (Lupo and Lewis 1979). The dotted curve allows for porosity in pure, cold, granular H$_2$O ice I$_h$, based on compaction data from Durham et al. (2005) as modeled by McKinnon et al. (2005). Higher densities are consistent with inclusion of materials denser than ice (such as silicates, carbonaceous species, or metals). Densities lower than the dotted curve are harder to interpret.

## 4. Conclusion

We present a technique for scheduling a sequence of astrometric observations to minimize the number required to characterize a transneptunian binary orbit. With Hubble Space Telescope, this method can often enable period, semimajor axis, and eccentricity of the mutual orbit to be determined with only four more observations, following the discovery observation. Observations scheduled according to this method reveal the mutual orbit of binary Centaur (42355) Typhon-Echidna to have semimajor axis $a = 1628 \pm 29$ km, period $P = 18.971 \pm 0.006$ days, and eccentricity $e = 0.526 \pm 0.015$. If Typhon and Echidna both share the same albedo, as implied by their similar colors, their radii are $76^{+14}_{-16}$ and $42^{+8}_{-9}$ km, respectively, based on radiometric observations by Spitzer Space Telescope. The average bulk density of the system is then only $0.44^{+0.44}_{-0.17}$ g cm$^{-3}$, consistent with strikingly low bulk densities recently reported for other small TNOs. This finding reinforces the impression of a compositional and/or thermal-history distinction between small TNOs with low densities and larger TNOs with much higher densities. Mutual events between Typhon and Echidna will be observable around 2019-2026, offering an additional opportunity for physical studies.

**Acknowledgments**

This work is based in part on NASA/ESA Hubble Space Telescope Cycle 14 program 10508 and 10514 observations. Support for these programs was provided by NASA through grants from the Space Telescope Science Institute (STScI), which is operated by the Association of Universities for Research in Astronomy, Inc., under NASA contract NAS 5-26555. We are especially grateful to Tony Roman at STScI for his quick action in scheduling HST follow-up observations. This work is also based in part on Spitzer Space Telescope program 55 observations. Spitzer is operated by the Jet Propulsion Laboratory, California Institute of Technology under a contract with NASA through an award issued by JPL/Caltech. We are grateful to D. Hestroffer for fruitful discussions and comparisons between his methods and ours and to two anonymous referees for their meticulous reviews. Finally, we thank the free and open source software communities for empowering us with key tools used to complete this project, notably Linux, the GNU tools, OpenOffice.org, MySQL, STSDAS, GDL, Python, FVWM, and TkRat.